\documentclass[conference]{IEEEtran}
\usepackage{cite}

\ifCLASSINFOpdf
\else
\fi
\usepackage{amsmath}

\usepackage{graphicx}

\begin{document}
%
\title{W-Net: A Two-Stage Convolutional Network for Nucleus Detection in Histopathology Image}

\author{
\IEEEauthorblockN{Anyu Mao$^{1}$, Jialun Wu$^{1}$, Xinrui Bao$^{2}$, Zeyu Gao$^{1}$, Tieliang Gong$^{1}$, and Chen Li$^{1,*}$}
\IEEEauthorblockA{$^{1}$School of Computer Science and Technology, Xi'an Jiaotong University, Xi'an, China\\
National Engineering Lab for Big Data Analytics, Xi'an Jiaotong University, Xi'an, China} 
\IEEEauthorblockA{$^{2}$School of Automation Science and Engineering, Xi'an Jiaotong University, Xi'an, China\\
Email: maoanyu@stu.xjtu.edu.cn}
}


%


\maketitle

\begin{abstract}
Pathological diagnosis is the gold standard for cancer diagnosis, but it is labor-intensive, in which tasks such as cell detection, classification, and counting are particularly prominent. A common solution for automating these tasks is using nucleus segmentation technology.
However, it is hard to train a robust nucleus segmentation model, due to several challenging problems, \textit{i.e.}, the nucleus adhesion, stacking, and excessive fusion with the background.
Recently, some researchers proposed a series of automatic nucleus segmentation methods based on point annotation, which can significant improve the model performance.
Nevertheless, the point annotation needs to be marked by experienced pathologists. 
In order to take advantage of segmentation methods based on point annotation, further alleviate the manual workload, and make cancer diagnosis more efficient and accurate, it is necessary to develop an automatic nucleus detection algorithm, which can automatically and efficiently locate the position of the nucleus in the pathological image and extract valuable information for pathologists.
In this paper, we propose a W-shaped network for automatic nucleus detection.
Different from the traditional U-Net based method, mapping the original pathology image to the target mask directly, our proposed method split the detection task into two sub-tasks. The first sub-task maps the original pathology image to the binary mask, then the binary mask is mapped to the density mask in the second sub-task. 
After the task is split, the task's difficulty is significantly reduced, and the network's overall performance is improved.
In order to evaluate the model efficiency in system application, we integrate the automatic nucleus detection model into the digital pathology image analysis system, Pathology Information Management \& Integration Platform (PIMIP). Our proposed network can automatic identify the center of each nucleus. Combined with the NuClick, a semi-supervised recognition model based on point annotation, we implement a fully automatic nucleus annotation framework.

\end{abstract}


%
\IEEEpeerreviewmaketitle

\section{Introduction}
Body fluid examination, imaging examination, and endoscopy can show the abnormalities inside the body, but whether the cancer is diagnosed must undergo pathological biopsy. Pathological diagnosis can further determine the cancer stage, and intuitively show the curative effect in the follow-up treatment process.

Pathological diagnosis is labor-intensive in which tasks such as nucleus counting and classification statistics are particularly prominent. These tasks need to determine whether the nucleus exists and where it exists, so it is necessary to do cell detection to locate the center of each nucleus, which is the basis for further pathological diagnosis.

The resolution of a complete digital pathological image is extremely large. In the image, the number of nuclei is enormous, but the shape of each nucleus is tiny. If we only rely on pathologists' manual annotation, it will be highly repetitive to pathologists and take up time and energy for follow-up observation and analysis of pathological tissues, and increase the pressure on pathologists.

There is a noticeable shortage of professional pathologists worldwide, which is not enough to support the pathological diagnosis needs of all humanity. Although the number of pathologists is increasing year by year, it still can not be proportional to the number of patients.

The automatic nucleus detection algorithm based on artificial intelligence can rely on the machine to complete the locating of the center of each nucleus in digital pathological images and further complete basic tasks such as counting nuclei. So it can replace the highly repetitive manual work and efficiently extract information valuable to doctors from digital pathological images. Pathologists can save more time and energy to analyze more detailed pathological information such as cell morphology.

In addition, recently, some researchers proposed a series of automatic nucleus segmentation methods based on point annotation, which can significant improve the model performance. But it requires more sophisticated nucleus detection, that is, locating the center of each nucleus.
Nevertheless, the point annotation needs to be marked by experienced pathologists. In order to take advantage of segmentation methods based on point annotation, further alleviate the manual workload, and make cancer diagnosis more efficient and accurate, it is necessary to develop an automatic nucleus detection algorithm, which can automatically and efficiently locate the center of each nucleus in pathological images for pathologists.


In pathology, there has been a method for coloring cells in tissues, that is Hematoxylin-Eosin Staining (H\&E). Current detection methods for nuclei in H\&E stained images are mostly based on low-level manual features, such as color, edge, and texture. Traditional algorithms or machine learning methods based on such features include local threshold \cite{oberlaender2009automated} and morphological calculation \cite{esmaeilsabzali2012machine} , region growing \cite{basavanhally2009computerized, ali2012integrated}, active contour model \cite{al2009active}, level set \cite{sethian1999level}, watershed algorithm \cite{kucharski2021cnn}, k-means clustering \cite{zhao2021k}, etc. However, for pathological images with numerous nuclei, the time cost of using the above methods is too high. For nuclei with irregular edges, methods based on edge features have limitations, while for images with complex backgrounds, methods based on color and texture have drawbacks.

With the rapid development of deep learning methods, a common solution for locating nucleus is using nucleus segmentation technology. The nucleus segmentation technology is challenging to train a robust model, and it is difficult to solve the problems of nucleus adhesion, stacking, and excessive fusion with the background. As for data preparation, the nucleus segmentation technology requires accurate nuclear boundaries, which requires high professionalism. However, the nucleus detection technology is easy to prepare a large amount of data, and it is easy to distinguish stacked nuclei by locating the center of each nucleus. Besides, automatic nucleus segmentation methods based on point annotation need nucleus detection to provide the center of each nucleus as necessary signals.

In order to solve the above problems, we have developed an automatic nucleus detection model based on deep learning and integrated it into the digital pathology analysis system, Pathology Information Management \& Integration Platform (PIMIP). Combined with the model named NuClick \cite{koohbanani2020nuclick} which is an automatic nucleus segmentation method based on point annotation, we build a complete automatic nucleus annotation framework. It can assist pathologists in their work, relieve their burden, and advance the digitization of pathology.

The main contributions of this paper are as follows:
\begin{itemize}
\item Design of nucleus detection model: We have made adaptive processing of the renal cell carcinoma data set. According to the characteristics of renal cancer cell data, we improved the U-Net based model specifically the feature extraction ability of the encoding part. We further split the detection task into two sub-tasks. The first sub-task maps the original pathology image to the binary mask, then the binary mask is mapped to the density mask in the second sub-task, so we have proposed W-Net.

\item System integrated development: In order to verify the actual effect of our model, we integrated the designed automatic nucleus detection model into the digital pathology analysis system PIMIP to provide a specific nuclear center for the automatic nucleus segmentation method based on point annotation and realize a fully automatic nucleus annotation framework.
\end{itemize}

\section{Related work}
In this chapter, from the perspective of network structure, we classify the existing automatic nucleus detection methods and analyze the advantages and disadvantages of various methods. Additionally, we introduce a pathological image analysis system PIMIP and the NuClick, a semi-supervised nucleus segmentation model based on point annotation, which is used in PIMIP.

\subsection{Automatic nucleus detection methods}
There are many nucleus detection methods based on deep learning including probability prediction models and image segmentation models. The above two types of methods both use Convolutional Neural Network(CNN) to extract the features of images, and then process the features to obtain final results. The subsequent operations on the features distinguish two types of network structures. One type is a network structure with only encoder, another is an encoder-decoder mode network. In addition, these two structures have a clear distinction between the form of the input and the efficiency of detection.

\subsubsection{Tiles analysis based on CNN downsampling}

This type of models includes Structured Regression CNN (SR-CNN) \cite{10.1007/978-3-319-24574-4_43}, Spatially Constrained CNN (SC-CNN) \cite{2016Locality}, Shape Prior CNN(SP-CNN) and Tunable Shape Prior CNN (TSP-CNN) \cite{2019Prior} , etc.

The structure of SR-CNN and SC-CNN is similar. They use CNN downsampling to obtain features and then connect to fully connected layers. The input required by the above two methods is a tile of a complete slide image which is cut by a sliding window with fixed size. SP-CNN and TSP-CNN are detection methods based on the prior shape of each nucleus, which introduce prior knowledge of the shape of each nucleus to do the detection. Although the input data required by these two methods in the detection part is a complete pathological image, the correction processing part requires a sliding window to extract tiles from the pathological image and compare it with prior knowledge. 

These methods predict the nucleus by predicting the probability of pixels. However, they adopt the form of a sliding window during data input, which is extremely inefficient for large-resolution digital pathological images. Among them, the setting of the sliding window's size needs to consider factors such as the size and number of nuclei, which requires professional opinions. The data preparation for prior information of the nuclear shape is difficult. In addition, the predicted value of any pixel will be affected by multiple blocks, which will have a negative impact on accuracy.

\subsubsection{Full-image prediction method based on encoder-decoder model}
This type of models includes Stacked Sparse Auto-encoder (SSAE) \cite{2014Stacked}, Fully Residual Convolutional Neural Network (FRCN) \cite{Weidi2018Microscopy} and U-Net \cite{2019U}.

SSAE is composed of multiple layers of basic auto-encoders. The features are normalized by softmax, and finally are restored to the original size through a symmetrical decoder.
In this method, whether the predicted value of the entire block is close to 1 is used to determine whether the block is a block with or without a nucleus. So it requires the input to be a picture extracted from a large-resolution pathological image, and there is a strict requirement that the block contains at most one nucleus. It makes the model training inefficient.

FRCN uses residual blocks for downsampling, and uses short-hop connections to avoid the problem of gradient disappearance, so that the model can learn features better.
FRCN and FRCN with structured regression \cite{2017Efficient} are both probability prediction methods. The output is predicted values of each pixel in the whole image. The higher the value, the closer to the center of each nucleus. So its truth ground is a density mask and input can be whole pathological images.
However, only when the distinction between nuclei and the background is obvious, the center of each nucleus can be highlighted. For pathological images with more complicated backgrounds, the performance of this model is poor.

In the medical image segmentation task, one of the best performing methods is U-Net \cite{2015U}, which uses an encoder-decoder structure and a long-hop connection between the encoder part and the decoder part, fusing low-level features with high-level features. The input can be a complex pathological image, and the output is a mask that distinguishes the target area from the background, which can also meet the requirements of nucleus detection.
U-Net can segment nuclei from images with complex backgrounds, but the drawback is that the center of each nucleus is not highlighted. Adhered or stacked nuclei may be mistakenly detected as the same nucleus.

In summary, the goal of our work is to directly predict the center of each nucleus from pathological images with complex backgrounds, so the encoder-decoder structure can meet our needs. We hope to take advantage of FRCN and U-Net, that is, nuclei can be detected from complex images, and the center of each nucleus can be highlighted to avoid problems such as missed detection and wrong detection.

\subsection{Pathology image analysis system}
PIMIP is a digital pathology image analysis system of our team. The goal of the system is to build a fully automatic digital pathology image annotation framework, but the existing technology still has limitations.

The system relies on NuClick \cite{2020NuClick}, a semi-supervised nucleus segmentation model based on point annotations to draw nuclear masks. Previous methods require a bounding box \cite{2017Interactive} or at least four annotations \cite{2017Extreme} as signals to segment nuclei and draw the mask. NuClick only needs one point annotation that is the center of each nucleus when segmenting the nucleus and drawing the corresponding mask.

The point annotation needs to be manually annotated in the system, which is a task with extremely high time cost and labor cost. In addition, there may be problems such as excessive position deviation and manual omission. In order for the framework of nucleus annotation to be more automatic, more efficient, and more accurate, it is necessary to convert manual annotations into automatic nucleus detection based on deep learning models.

The automatic nucleus detection model designed in this paper should be integrated into PIMIP, which returns the coordinates of nuclear centers, and provides it to NuClick model as guide signals to draw nuclear masks. Finally, we can build a complete automatic nuclear annotation framework.

\section{Methods}
We design the model based on the encoder-decoder structure. We improve the U-Net based model specifically the feature extraction ability of the encoding part from a multi-dimensional perspective. In order to reduce the difficulty of the detection task and make the result more accurate, we split the task into two sub-tasks by adding an intermediate prediction step. We propose W-shaped network(W-Net) to improve the overall performance of the model.

\begin{figure*}
    \centering
    \includegraphics[width=150mm]{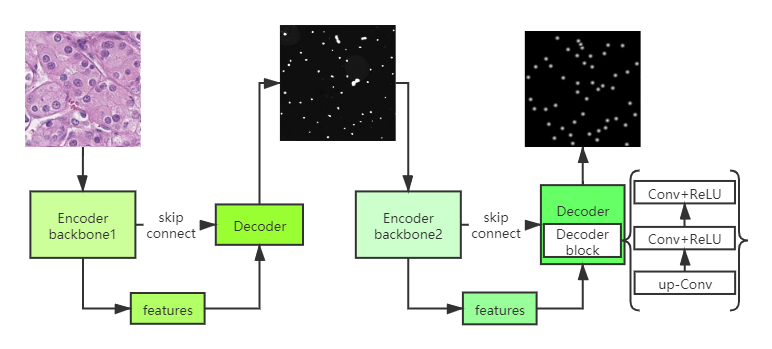}
    \caption{W-Net. Our proposed method W-Net splits the detection task into two sub-tasks. The first sub-task maps the original pathology image to the binary mask, then the binary mask is mapped to the density mask in the second sub-task. In two sub-tasks, we both use encoder-decoder structure. For the encoder of the first sub-task, we use EfficientNet-B7 as its backbone. When extracting the features of the binary mask, the background has been distinguished from the nucleus, so a simpler structure of encoder can be used. We finally choose EfficientNet-B0 for the second sub-task.}
    \label{f1}
\end{figure*}

\subsection{Improve the encoder of U-Net}
U-Net can be applied to nucleus detection. We only need to re-plan the features we want the network to extract. In short, it is to change the form of the label to meet the goals. However, the encoder used by the U-Net is simple. Due to the limitation of the depth, the degree of extracted features is limited, resulting in low recall rate. Many non-existent nuclei will be detected. In addition, the simple encoder cannot eliminate the negative impact caused by adhered and stacked nuclei.

Therefore, we improve the encoder by using a pre-trained network as its backbone. At the same time, the decoder part is modified to restore the features to images with original size based on the principle of symmetry. The basic decoding block is a deconvolution layer connecting two convolution layers and an activation layer, which is shown in Fig. 1. The number of decoding blocks is adjusted according to the encoder to restore the size of the images symmetrically. The deconvolution mode is bilinear interpolation.
Long-hop connections are still used to integrate low-level features and high-level features. The following are two frameworks used as the encoder backbone.

\subsubsection{Deep Residual Network (ResNet)}
The depth of the network is crucial to the performance of the model. When the number of network layers is increased, the network can extract more complex features and theoretically achieve better results. But the deep network will have degradation problems. The residual block can effectively solve the above problems. Therefore, the U-Net with ResNet \cite{2016Deep} is better in performance of detection. There are experiments using ResNet-34 as the U-Net encoder backbone, while 50, 101, and 152 are not widely used, so we focus on these four kinds of ResNet to find the most suitable number of layers.

\subsubsection{EfficientNet}
ResNet improves accuracy by increasing the depth of the network, but it affects the speed of model. Besides, only considering the depth of the network is too simple, which will cause the improvement of network performance to fall into a bottleneck. EfficientNet explores a more suitable combination of network depth, width and image resolution, and has a good trade-off between speed and accuracy \cite{DBLP:journals/corr/abs-1905-11946}. Compared with ResNet, EfficientNet not only improves the speed and reduces the amount of parameters to a certain extent, but also maintains accuracy and even improves the performance of the network when the amount of calculation is equivalent.

Therefore, we use EfficientNet as the encoder backbone to construct a U-Net with EfficientNet. We try EfficientNet-B5, B6, and B7 which is different in the zoom levels in multiple dimensions to find out the most suitable structure.

\subsection{W-Net}
U-Net with EfficientNet mentioned above greatly improves the ability of feature extraction of the encoder, which reduces the problems of over detection and error detection. The predicted results basically distinguish each nucleus, and the nuclei with a lighter degree of adhesion can be distinguished. But the stacked nuclei are not distinguished. Areas where nuclear color is not highly distinguished from background's color are even more difficult to highlight the center of each nucleus.

Analyzing the characteristics of inputs, it can be found that the background of some images is similar to the color of the nucleus. As shown in Fig.2 (a), some of nuclei can almost blend with the background. Once the stacked nuclei are fused with the background, it is difficult to determine where the highest value of the nucleus center pixel starts. Therefore, it is very difficult to determine the center coordinates in such case. 

So the method of directly predicting centers of each nuclei from the pathological image with complex backgrounds makes detection task difficult. Therefore, we split the detection task into two sub-tasks. The first sub-task maps the original pathology image to the binary mask that distinguishes the background and nuclei, then the binary mask is mapped to the density mask in the second sub-task that highlights the center of each nucleus. In this way, when predicting the center of each nucleus, the stacked nuclei will not be affected by excessive fusion with the background, and the task of highlighting the center of each nucleus is greatly simplified.

\begin{figure*}
    \centering
    \includegraphics[width=120mm]{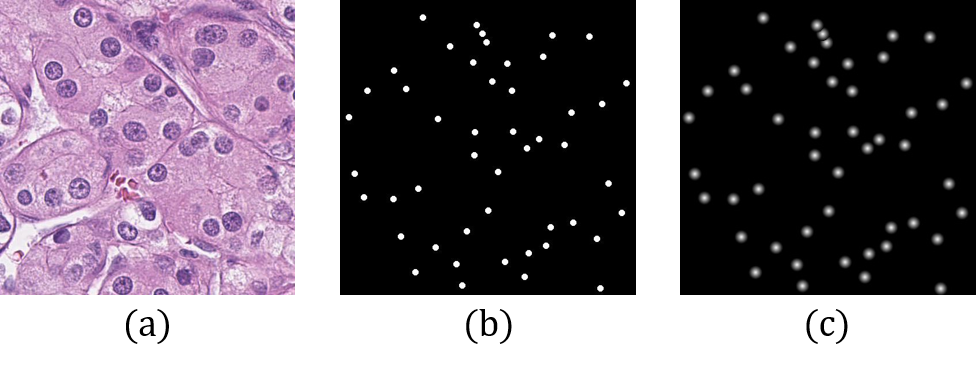}
    \caption{The result of the data processing. (a) Original image. (b) Solid dots. (c) Density mask.}
    \label{f2}
\end{figure*}

In summary, our proposed method W-Net splits the detection task into two sub-tasks. After the task is split, the task's difficulty is significantly reduced, and the network's overall performance is improved. The W-Net architecture is shown in Fig. 1.

In two sub-tasks, we both use encoder-decoder structure. When extracting features from original images, the encoder with better ability of feature extraction is used. We find that the U-Net with EfficientNet has better performance. Therefore, for the encoder of the first sub-task, we use EfficientNet as its backbone. When extracting the features of the binary mask, the background has been distinguished from nuclei, so a simpler structure of encoder can be used. Different choices and comparisons of encoders for the second sub-task are introduced as follows. 

\subsubsection{Basic encoder}
U-Net cannot distinguish the adhered or stacked nuclei, but it can perform better in binary masks that distinguish the background and nuclei, and highlight the center of each nucleus.

\subsubsection{VGGNet}
Since the basic encoder does not have pre-trained parameters, there is randomness in the training process, and there are still limitations for extracting the features. Among the pre-trained models with a small number of layers, there is VGGNet \cite{2014Very} that is commonly used to extract image features.
VGGNet has a simple structure and uses a small convolution kernel instead of a large convolution kernel. A small convolution kernel means more activation functions, richer features, and stronger discrimination capabilities. The small convolution kernel also reduces the parameters of the convolution layer. Therefore, in the second sub-task, we use VGGNet to improve the performance of network. VGGNet-13 and VGGNet-16 are tried as the backbone of the encoder for the second sub-task.

\subsubsection{EfficientNet with different zoom levels}
EfficientNet comprehensively considers the performance of the network from the three dimensions of network depth, network width, and image resolution. U-Net with EfficientNet performs well in detection. Binary masks eliminate the negative impact of excessive fusion of the background and nuclei, which greatly reduce the difficulty of predicting the center of each nucleus. It will better utilize EfficientNet's ability. We use EfficientNet-B0 as the backbone of the encoder for the second sub-task.

\section{Result}
We compare the performance between U-Net, U-Net with ResNet, U-Net with EfficientNet and W-Net. The data set used is annotated by our team.

\subsection{Renal clear cell carcinoma data set}
The data set used is named clear cell Renal Cell Carcinoma Nuclei Grading Data set (ccRCC), which is composed of 1000 H\&E stained tiles with a resolution of 512 pixels × 512 pixels.
The labels in the ccRCC contain the coordinates of centers of nuclei. There is a problem that if the label is only a single-point pixel in the center of the nucleus, it will make the convolutional network unable to extract features. Therefore, it is necessary to process the data. The specific methods are as follows.

\subsubsection{Solid dots}
The center of each nucleus occupies one pixel, and it is marked as 1. The proportion of single-point pixels is too small compared to the proportion of background pixels marked as 0, which makes the encoder difficult to extract features that the center of each nucleus is marked as 1. The output is close to a whole black image and nuclei are almost undetectable.

Therefore, we expand the influence of the center of each nucleus and increase the proportion of pixels marked as 1. We consider the shape of a dot as the prior shape of the center of each nucleus, and describe the center of each nucleus as a dot with a radius of 2 pixels, which is shown in Fig.2 (b).

\begin{table*}
\caption{The comparison of U-Net with different encoders}
\begin{center}
\begin{tabular}{|c|c|c|c|}
\hline
Methods         & P    & R    & F1   \\ \hline
U-Net           & 0.78 & 0.80 & 0.78 \\ \hline
U-Res-Net       & 0.81 & 0.84 & 0.82 \\ \hline
U-E-Net         & \textbf{0.82} & \textbf{0.87} & \textbf{0.84} \\ \hline
\end{tabular}
\end{center}
\end{table*}

\begin{table*}
\caption{The comparison of U-Net and W-Net }
\begin{center}
\begin{tabular}{|c|c|c|c|}
\hline
Methods         & P    & R    & F1   \\ \hline
U-Net           & 0.78 & 0.80 & 0.78 \\ \hline
W-UU-Net        & 0.81 & 0.79 & 0.80 \\ \hline
W-EU-Net        & 0.82 & 0.84 & 0.83 \\ \hline
W-EV-Net        & 0.82 & 0.86 & 0.84 \\ \hline
Proposed Method & \textbf{0.83} & \textbf{0.88} & \textbf{0.85} \\ \hline
\end{tabular}
\end{center}
\end{table*}

\subsubsection{Density mask}
Although the solid dots mask enlarges the influence of the center of each nucleus, it also spreads the center of the nucleus, leading to the problem of predicted center shift. Therefore, we need to highlight the center points. The labels of breast cancer nucleus data set (BCData) \cite{2020BCData} are generated based on the method of the crowd counting model CSRNet \cite{2018CSRNet}. It uses Gaussian kernel convolution, so that the central pixel points are spread around, but maintain highest.
However, the pixel values of the image become extremely small, resulting in the network not being able to extract features. Therefore, considering the density of nuclei, the specific range of diffusion is selected. According to the density mask generation method in FRCN with structured regression, the center point keeps value 1, and it spreads into a circle with a radius of d pixels, and the ratio of the decrease in pixel value is related to distance to the center. Suppose the pixel value of a point is $M_{ij}$, the distance that the point to the center is $D$, and $\alpha$ is a given ratio. The formula is shown as (1), and the density mask generated according to the formula (1) is shown as Fig. 2 (c).

\begin{equation}
M_{ij}=\begin{cases} \frac{ e^{\alpha \left ( 1-\frac{D\left ( i,j \right )}{d} \right )-1}}{e^{\alpha }-1} & \text{ if } D\left ( i,j \right )\leq d \\ 
0 & \text{ otherwise} 
\end{cases}
\end{equation}

We divide the data set into three subsets, including 600 tiles in the training set, 200 tiles in the validation set, and 200 tiles in the test set.

\subsection{Experiment details}
For the evaluation of the model, we choose F1 score. F1 score is the harmonic average of precision (P) and recall (R). The maximum value is 1 and the minimum value is 0. The higher the value is, the better the model is. The formula is shown in (2).
\begin{equation}
    F_{1}=2\times \frac{P\times R}{P+R}
\end{equation}

True Positive (TP) indicates that the distance between the predicted center of the nucleus and true center is less than 5 pixels.
False Positive (FP) indicates that the predicted point is not within the allowable range of the true center, that is, the distance from the true center exceeds 5 pixels.
False Negative (FN) indicates the point that is not predicted among the truth points. Therefore, the calculation formulas for P and R are shown in (3) and (4).
\begin{equation}
    P= \frac{TP}{TP+FR}
\end{equation}
\begin{equation}
    R= \frac{TP}{TP+FN}
\end{equation}

We used a GPU server to train the network model, which is configured as an RTX 2080Ti graphics card, with the storage up to 11000MiB.
After training every 10\% of the training samples, we verify the current network to adjust the learning rate. The initial learning rate is set to 0.0001, and the minimum learning rate is set to 0.00000001.

The loss function used for model training includes the cross-entropy loss function for two classifications (BCE-Loss) and the average absolute error loss function (L1-Loss). In the comparison experiment of U-Net with different encoders, we mainly use L1-Loss. In the comparison experiment of W-Net, the first sub-task is regarded as a binary classification problem with pixels other than 0 or 1, so BCE-Loss is used. And the second sub-task is regarded as the regression problem of the two-dimensional matrix to the density mask, so the L1-Loss is used. In the W-Net training process, the loss calculated by the two loss functions is quite different in magnitude. In order to prevent any loss from being ignored, the ratio $\alpha$ of the two is adjusted. The loss formula for W-Net is shown in (5).
\begin{equation}
    Loss_{W-Net}=Loss_{BCE}+\alpha\times Loss_{L1}
\end{equation}
\subsection{Experiment result}
We compare the performance between U-Net, U-Net with ResNet, U-Net with EfficientNet and W-Net.

\begin{figure*}
    \centering
    \includegraphics[width=180mm]{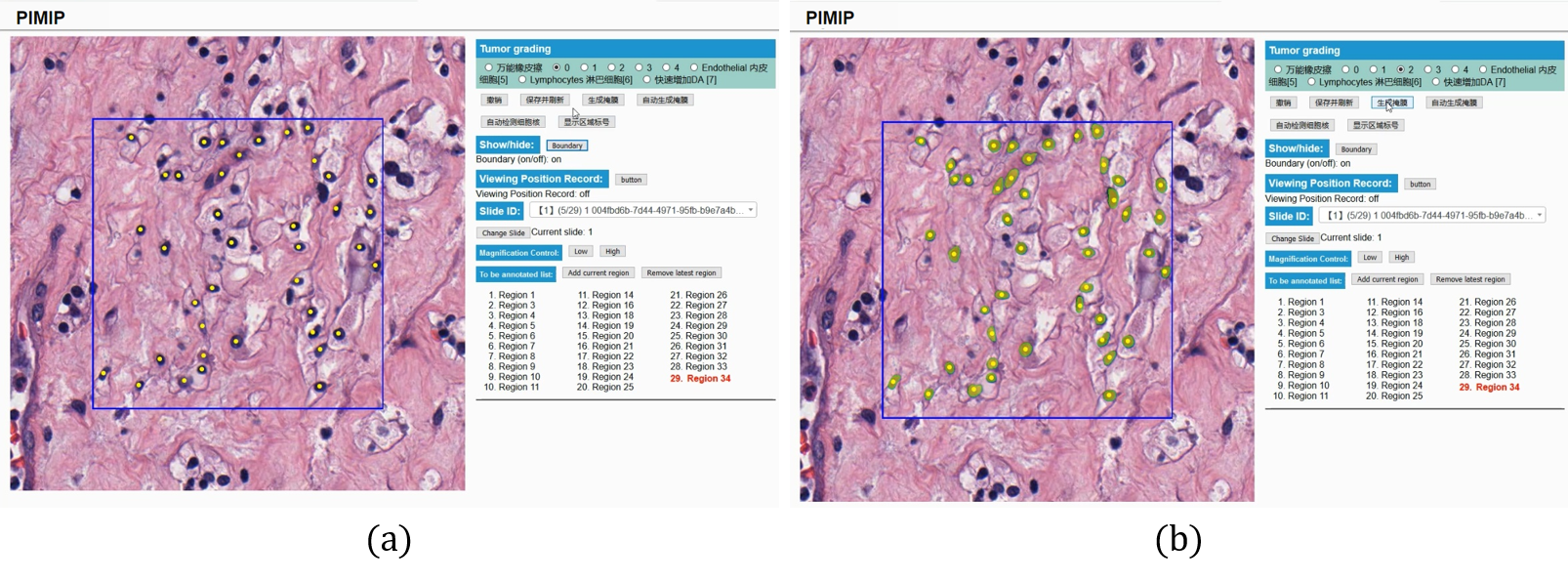}
    \caption{The effect of PIMIP's nucleus annotation. (a) The result of automatic nucleus detection using W-Net. (b) After transmitting the guiding signal of the nucleus center gained by W-Net, the NuClick model is used to complete the automatic nuclear mask drawing. We construct a complete automatic nucleus annotation framework.}
    \label{f3}
\end{figure*}
 
The comparison of U-Net with different encoders are shown in table 1. Experiments are carried out on U-Net, U-Net with ResNet-34, 50, 101 and 152 as encoders and U-Net with EfficientNet-b5, B6 and B7 as encoders. The experimental data of U-Net is used as the baseline to show the effect of improving encoders. It is found that when the encoder takes ResNet as the backbone, the effect of ResNet-50 is the best. When EfficientNet is used as the encoder backbone, EfficientNet-b7 is the best choice. U-Res-Net refers to U-Net with ResNet-50 as its encoder backbone and U-E-Net refers to U-Net with EfficientNet-B7 as its encoder backbone. 

The comparison of U-Net and W-Net are shown in table 2. According to the above group of comparative experiments, it is determined that the first sub-task still uses EfficientNet-B7 as the encoder backbone. For the encoder backbone of the second sub-task, we try the basic encoding blocks, VGGNet-13, 16 and EfficientNet-B0 respectively. When trying VGGNet, it is found that VGGNet-16 works best. W-UU-Net means W-Net consisting of two U-Nets. W-EU-Net means that EfficientNet-B7 is used as the first sub-task's encoder backbone and the basic encoder is used in the second sub-task. W-EV-Net means that EfficientNet-B7 is used as the first sub-task's encoder backbone and VGGNet-16 is used as the second sub-task's encoder backbone. The method we proposed, W-EE-Net, uses EfficientNet-B7 as the first sub-task's encoder backbone and EfficientNet-B0 as the second sub-task's encoder backbone.

The effect of the basic encoder is not as good as using pre-trained models. Due to the insufficient downsampling depth of the network, the network can not extract better features, resulting in poor prediction results. When using ResNet and EfficientNet with deeper network depth as encoder backbones, the features extracted by the network are more conducive to the correct prediction of the nuclear center. U-Net with EfficientNet performs better than W-Net with ResNet in recall when the precision is improved. The reason is that ResNet only considers the network depth to improve the network performance, while EfficientNet considers network performance from the three dimensions of depth, width, and image resolution.

The W-Net we proposed significantly surpasses U-Net in terms of precision and recall rate, and compares with the best performing U-E-Net, the recall rate has also been improved, making the overall performance of the model promoted. The task is split, which reduces the difficulty of the task and gives better play to the encoding ability. The U-Net needs to learn from pathological images with a complex background to highlight the center of each nucleus. That makes it too difficult to distinguish between the background and nuclei with a high degree of fusion in a single encoding and decoding process. W-Net firstly distinguishes the background and nuclei, solving the problem of fusion of nuclei and the background, and then map the binary mask to the density mask to highlight the center of each nucleus. After excluding the influence of graph fusion, the difficulty of the task is reduced, so the performance of the network is improved apparently.

\subsection{System integration development}
We integrated the designed W-Net into PIMIP.
Firstly, the automatic nucleus detection function is realized according to the MVC framework. A new button for auto-detecting nuclei is added to the user interface, which triggers the back-end model prediction function through the parameter transmission of the controller. The prediction function stores the predicted coordinate points in the server so that the controller can read and display it on the front-end interface. The user interface also includes a manual correction function to manually add or delete nuclei.

We connect the newly added automatic nucleus detection function with the NuClick model. By transmitting the guiding signal of the nucleus center, the NuClick model is used to complete the automatic nuclear mask drawing. We construct a complete automatic nucleus annotation framework.

The result of PIMIP's nucleus annotation is shown in Fig. 3. The performance of the model in the system is good, and the system integration development is successful.

\section{Conclusion and Discussion}
\subsection{The Conclusion}
In the process of pathological diagnosis, there is a large part of manual work with a high degree of repetition, especially the work of annotating nuclei. Moreover, the nucleus segmentation technology is hard to train a robust model, and it is difficult to solve the problem of nucleus adhesion, stacking or excessive fusion with the background.

In order to solve the above problems, we develop an automatic nucleus detection algorithm based on deep learning, train a robust model, and integrate it into PIMIP system to build a complete automatic nucleus annotation framework to reduce the burden of pathologists.

We have made adaptive improvements to the network with encoder-decoder structure, experimented with basic encoding blocks, U-Net with ResNet, and innovatively tried the U-Net with EfficientNet. It is found that increasing the depth of the network, and considering the depth, width and image resolution comprehensively, is conducive to improving the network's ability to extract image features and improving the performance of the network in nucleus detection.

By analyzing the prediction results of the U-Net, we finally split the detection task into two sub-tasks and add an intermediate step to form a learning path which maps the original image to the binary mask that distinguishes the background and nuclei and then maps to the density mask that highlights the centers of nuclei. So that W-Net is proposed, in which EfficientNet-B7 is the backbone of the encoder for first sub-task, and  EfficientNet-B0 is the backbone of the encoder for second sub-task. The experiments found that after the task was split, the difficulty of the task was greatly reduced, and the overall performance of the model was improved.

We also have carried out system integration development. We integrated W-Net into PIMIP system, and combined the detection function with the NuClick model to realize full automatic nucleus annotation framework.

\subsection{The Discussion}
In the process of improving the encoder of the U-Net, two types of backbones have been tried. In the future, more innovative and better encoder backbones should be tested. Improving the decoder is also important for promoting the performance of the model.

Although W-Net shows better performance than the U-Net in predictions, the speed is significantly reduced and the model occupies more storage space. The problem of storage and speed also leads to great limitations in the choice of the encoder structure. Therefore, the model will be pruned in the follow-up to reduce unnecessary parameters and improve the speed of training and prediction.

For the system development, the first problem is that the annotation interface lacks more detailed guidance information. The manual correction function needs to update in the real-time performance. Automatic nucleus detection models have different features such as high speed or high accuracy. However, the choice of models is only for code writers, and users lack independent choice. Users should have the right to choose the appropriate model according to their own needs.

All in all, in terms of the model design, we hope to increase the speed of model training and prediction, reduce model parameters, and try more encoders with different structures. After the model is further improved, a better performance and more robust model is trained to be used for automatic nucleus detection. The automatic nucleus detection function in the PIMIP system is currently working well, but we hope to further improve the real-time performance and user autonomy.


\section*{Acknowledgment}
This work has been supported by National Natural Science Foundation of China (61772409); The consulting research project of the Chinese Academy of Engineering (The Online and Offline Mixed Educational Service System for “The Belt and Road” Training in MOOC China); Project of China Knowledge Centre for Engineering Science and Technology; The innovation team from the Ministry of Education (IRT\_17R86); and the Innovative Research Group of the National Natural Science Foundation of China (61721002). The results shown here are in whole or part based upon data generated by the TCGA Research Network: https://www.cancer.gov/tcga.



%


\end{document}